\documentclass{aastex631}
\usepackage{csquotes}
\usepackage[hyphens]{url}
\usepackage{xurl}

\shorttitle{SciWri GE Courses}
\graphicspath{{./}{figures/}}
\begin{document}

\title{Effects of Popular Science Writing Instruction on General Education Student Attitudes Towards Science: A Case Study in Astronomy}

\correspondingauthor{Briley Lewis}
\email{blewis@astro.ucla.edu}

\author[0000-0002-8984-4319]{Briley L. Lewis}
\affiliation{Department of Physics and Astronomy, UCLA, Los Angeles, CA 90095 USA}
\affiliation{Cluster Program, UCLA, Los Angeles, CA 90095 USA}

\author[0000-0001-8630-8839]{K. Supriya}
\affiliation{Center for Education Innovation and Learning in the Sciences (CEILS), UCLA, Los Angeles, CA 90095 USA}
\affiliation{Center for the Integration of Research Teaching and Learning (CIRTL), UCLA, Los Angeles, CA 90095 USA}

\author[0000-0002-7716-6506]{Graham H. Read}
\affiliation{Molecular Biology Institute, UCLA, Los Angeles, CA 90095 USA}
\affiliation{Center for the Integration of Research Teaching and Learning (CIRTL), UCLA, Los Angeles, CA 90095 USA}

\author[0000-0001-6821-357X]{Kaitlin L. Ingraham Dixie}
\affiliation{Center for Education Innovation and Learning in the Sciences (CEILS), UCLA, Los Angeles, CA 90095 USA}
\affiliation{Center for the Integration of Research Teaching and Learning (CIRTL), UCLA, Los Angeles, CA 90095 USA}

\author[0000-0001-8461-2204]{Rachel Kennison}
\affiliation{Center for Education Innovation and Learning in the Sciences (CEILS), UCLA, Los Angeles, CA 90095 USA}
\affiliation{Center for the Integration of Research Teaching and Learning (CIRTL), UCLA, Los Angeles, CA 90095 USA}

\author[0000-0003-1249-2041]{Anthony R. Friscia}
\affiliation{Department of Integrative Biology and Physiology, UCLA, Los Angeles, CA 90095 USA}
\affiliation{Cluster Program, UCLA, Los Angeles, CA 90095 USA}



\begin{abstract}
For many students, introductory college science courses are often the only opportunity in their formal higher education to be exposed to science, shaping their view of the subject, their scientific literacy, and their attitudes towards their own ability in Science, Technology, Engineering, and Math (STEM). While science writing instruction has been demonstrated to impact attitudes and outlooks of STEM majors in their coursework, this instructional strategy has yet to be explored for non-majors. In this work, we investigate student attitudes towards STEM before and after taking a writing-intensive introductory astronomy course. We find that students cite writing about science as beneficial to their learning, deepening their understanding of science topics and their perspective on science as a field and finding writing to be a ``bridge'' between STEM content and their focus on humanities in their majors. Students also report increased perceptions of their own ability and confidence in engaging with STEM across multiple metrics, leaving the course more prepared to be informed, engaged, and science literate citizens.
\end{abstract}



\section{Introduction} \label{sec:intro}

In the past few years, disbelief in science has been an issue of national news, especially during the COVID-19 pandemic, and the general public's attitudes towards science are often filled with misconceptions, leading to polarization in beliefs and mistrust of scientists \citep{rutjens2018attitudes,kreps2020model}. Yet, at the same time, science is respected as a career and often seen as requiring innate talent and intelligence, especially in the physical sciences; there are many inaccurate stereotypes of scientists that are pervasive in society as well, such as the ``lone male white genius'' trope \citep{bruun2018identifying}. These negative attitudes and misconceptions about Science, Technology, Engineering, and Math (STEM) fields naturally lead us to ask -- how do students develop these beliefs? How has their science education failed to address these beliefs, and what can we do to make a positive impact on students' attitudes towards science? 

For many students, their main exposure to science in higher education is an introductory course that fulfills a general education (GE) requirement. Research has shown that interventions to teach scientific literacy in science courses for non-science majors (e.g. introductory GE courses) can be quite effective in their ability to engender positive attitudes towards science and build science literacy skills \citep{wittman2009shaping, hobson2008surprising}. Links between increased science literacy and improved attitudes towards science in introductory astronomy courses for non-majors \citep{wittman2009shaping,duncan2012improving} have been well-established in the literature and provide support that attitudes and beliefs about science can be altered by effectively teaching science literacy.

Interventions involving current research and science writing are well established in the literature for STEM students, but there is less information available for these techniques in an introductory non-major/GE course. Studies show that engaging with recent science literature exposes STEM students to active research questions and provides real-world motivation to learn the content, making it clear how the course is relevant to their careers \citep{donohue2021integrating}. More generally, research has shown that providing students with a real-world writing task is incredibly impactful, leading to changes in student beliefs, and deepening student learning \citep{kiefer2008client}. Additionally, STEM students benefit from broader perspectives that can be developed by targeted interventions in discipline-based reading and writing skills. These writing skills help students develop broader perspectives on their work in science, put science concepts into context, and learn how to contribute to scientific discourse \citep{sorvik2015scientific, szymanski2014instructor, lewisprepWAC}. Writing is a prime example of an active learning strategy, which has been shown to be pedagogically inclusive in many disciplines \citep{penner2018building, theobald2020active}. Science writing has also been shown in the literature to be an effective tool for improving attitudes towards science in STEM students, but this result has not yet been extended to non-major/GE students \citep{pelger2016popular,erkol2010effect}. 

The goal of this project is to demonstrate that popular science writing can be an effective tool for improving student attitudes towards science and increasing student engagement with science in an introductory general education science course. That is, students in such a course generally do \textit{not} intend to major in a STEM discipline, and are taking the course to fulfill a science requirement amidst their major coursework in the humanities. In this work, we report qualitative information gathered on student experiences and attitudes in a uniquely writing-focused introductory GE astronomy course, investigating how science writing interventions with non-STEM students affected multiple factors: students' interest in science/astronomy, their perception of the benefits and personal relevance of science, their perception of their own ability to engage with and communicate about STEM, and their perception of stereotypes of scientists such as STEM fields requiring innate talent.

%We hypothesize that a science writing intervention in this introductory GE astronomy course will have a positive effect on non-major students' perceptions of science. Based on experience with a prior iteration of the course in this study, we believe that science writing instruction will help students of diverse non-STEM majors use writing as a method to connect science concepts to their own interests and their humanities coursework, and that writing as a real-world task will engender a belief that students are capable of engaging with scientific concepts.

The factors measured within this work all have a basis in the literature as desirable outcomes for students in an introductory non-major STEM course, or as factors that improve student learning. Student interest in a subject has been linked to student engagement and learning in a course -- naturally, if students are interested and curious about a topic, they are more likely to apply themselves to learning about it \citep{renninger2015power}. Thus, one measure of a "successful" introductory science course is an increase in student interest in the subject. Additionally, increased student understanding of the benefits of learning the subject matter at hand and of the relevance of the material to their lives is key for increasing student buy-in \citep{cavanagh2016student,wang2021framework,stuckey2013meaning, newton1988relevance,frymier1995s}.

When discussing student attitudes about science, there are three main categories of thought about STEM fields we want to influence: what they think of \textit{scientists}, how students think of the \textit{benefits} of learning science, and how they think of the \textit{relevance} of science to their lives / society. Many introductory GE science courses aim to promote diverse representation of current STEM practitioners and positively influence the views of students about scientists \citep{yuretich2001active}. This is in response to persistent stereotypes of scientists, especially for the physical sciences like physics and astronomy. Stereotypes regarding gender, age, ethnicity, and more exist, such as the idea that stereotypically female gender roles aren't compatible with the skills and personality of a scientist \citep{carli2016stereotypes}, that scientists are generally older white men \citep{ferguson2020exploring, bruun2018identifying, losh2010stereotypes}, and that being a scientist requires a high degree of innate talent \citep{leslie2015expectations}. The benefits of science are not only important for student buy-in, but also public buy-in to public funding for science \citep{munoz2012willing}. Relevance is also often cited as a key metric of good science teaching \citep{newton1988relevance} -- a 2013 review even described that the literature states ``making science learning relevant both to the learner personally and to the society in which he or she lives should be one of the key goals of science education'' \citep{stuckey2013meaning}. As perceptions of science courses became more negative in the 2000s, studies cited ``irrelevance'' as a student concern about their science education, and as a result there was a large push to connect science to students' lived experiences and better motivate them to learn about science \citep{stuckey2013meaning, holbrook2003increasing,holbrook2005making,gilbert2006nature, dillon201617}. Although relevance is often ill-defined in education studies, \citet{stuckey2013meaning} suggests three dimensions: relevance in preparing students for careers, relevance for understanding scientific phenomena, and relevance for students becoming effective future citizens. In this study, we asked students about the broad concept of relevance, to which students responded with their personal interpretations.

We hypothesize that a science writing intervention in this introductory GE astronomy course will have a positive effect on non-major students' perceptions of science. Based on experience with a prior iteration of the course in this study, we believe that science writing instruction will help students of diverse non-STEM majors use writing as a method to connect science concepts to their own interests and their humanities coursework, and that writing as a real-world task will engender a belief that students are capable of engaging with scientific concepts.

There are also four tasks that we investigate to judge student ability/confidence in science-related activities, all related to activities they have practiced during the course: reading primary literature, reading science news, explaining science concepts, and writing about a scientific discovery. High amounts of technical jargon often make it difficult for students to read primary literature in the science, with even graduate students reporting apprehension reading research articles \citep{lewisprepWAC}. Yet, despite research literature's inaccessibility, K-12 and undergraduate classes seek to incorporate cutting-edge research and even primary sources into their curriculum to increase relevancy and real-world applicability \citep{nieves2020bitescis,eales2016establishing}. Understanding the role of scientific publishing, and its reliability as a source, is also key to public perceptions of science \citep{miller2004public}. Additionally, in science writing and journalism, tackling research papers is one of the first -- and often most challenging -- steps in the writing process that a new writer has to learn \citep{witze_2020}. There are clearly many benefits to exposing students to primary sources in scientific research, but there is also the distinct challenge of approaching a genre written for an expressly technical audience. This task is expected to be the most difficult and intimidating, followed by writing about a discovery (which requires significant knowledge and synthesis), while reading science news and informally explaining concepts are expected to be more accessible to students, and more likely to be activities they continue in their daily lives.

In Section \ref{sec:methods}, we detail the study population as well as the design of the course and assessments. In Section \ref{sec:results}, we report and explore student responses to open-ended survey questions both before and after the course. Finally, in Section \ref{sec:disc}, we explore the implications of these results and offer recommendations for writing-based science education and future work on this topic.

\section{Methods} \label{sec:methods}

\subsection{Course Description and Study Population}

Student participants were all first-year undergraduates at a large public university in the United States at the time of the study. These students were enrolled in a seminar on astrobiology and science journalism. This seminar is part of a program offered only to first-year students that provides a range of year-long (three academic quarters) courses that cover ``Big Ideas,'' and all fulfill various general education (GE) requirements. Each year, this program offers ten of these multi-disciplinary courses, drawing faculty and teaching assistants from multiple departments. These courses are open only to incoming freshman students, and about one-third of the university's freshman class enrolls in a course through this program.

The first two quarters of these courses are ``traditional'' lecture/discussion section courses -- students attend lecture with the entire class two or three times a week and a smaller discussion section of 20 students led by a Teaching Assistant once a week. During the third quarter of the course, students choose a seminar to take, taught by a member of the teaching team which delves deeper into a related topic in a small (20 student) seminar course.  Specifically, student participants in this study were enrolled in the year-long program focusing on the evolution of life and the Universe.  The teaching team for this course includes an astronomer, geologist, and two biologists, and covers the history of the universe from the Big Bang to the evolution of humanity, including topics on the formation of the elements, planetary geology and biological evolution.  The emphasis is on science as a process rather than specifics of a particular field or topic. This series of courses fulfills their GE requirement for scientific inquiry, as well as their disciplinary writing requirement. 

The disciplinary writing requirement is satisfied by courses where students learn to write within a specific discipline/field.  In this year-long program, students write short (5-6 page) papers in the first two quarters of the course, and also receive instruction on using the library and how to read scientific literature. The assignment in the first quarter is a literature review based around one of a list of primary scientific papers provided to the students. They must put the paper in the broader context of its field, understanding how it builds on prior knowledge and is in turn built upon. For the assignment in the second quarter, students chose from a provided list of scientific debates and must use the skills learned in the previous quarter to research the debate and come to their own conclusion about the most supported viewpoint.  With both papers, the assignment is scaffolded, with various parts due throughout the quarter and students are given feedback throughout the writing process. The bulk of writing expected to satisfy the requirement, though, is meant to be part of the third quarter seminar course.

The course in question for this study is a seminar in the third and final quarter of the year-long program, focusing on astrobiology and science journalism. It is designed as a writing-intensive science course for non-STEM majors. Since this course fulfills science requirements, students with non-STEM majors typically enroll. Additionally, since this year-long experience is not a required program, there is a possibility that the program self-selects for particularly interested and motivated students, and students have some prior exposure to concepts about science from the first two quarters -- to adjust for the latter, we compare pre- and post-course surveys to get a sense of the changes that happened over the duration of only the writing-intensive science course, as opposed to the whole year-long program. 11 out of 21 students from the 2021 offering of the course consented to the retroactive use of information from their course evaluations for this study, and 19 out of 21 students from the 2022 course responded to the assessments designed for this study. 5 students from those two course offerings agreed to interviews to discuss their experiences. Demographic information such as race and gender was not collected from participants, as it is outside the scope of this study.

\subsection{Interventions}

In this writing-intensive science course, students worked through projects to develop a popular science article, from idea to pitch to full draft, on some topic related to astrobiology (e.g. planetary science, search for extraterrestrial intelligence (SETI), astronomy, space exploration). They had two major writing assignments: (1) a “translation” piece where they summarize a research article for a lay audience using the Astrobites format \citep{sanders2012preparing,sanders2017incorporating,khullar2019astrobites,lewisprepastrobites} and (2) a more open-ended project that is scaffolded to take them from an initial idea to a fully developed article with an accompanying pitch for publication.

These major assignments were supported with smaller assignments to explore other multimedia forms of science communication (e.g. social media, podcasting, video), a weekly writing journal to practice science writing skills, and readings from science publications, both technical and non-technical. There were also in-class activities to build various writing skills, such as genre awareness, knowledge of audience, skill in word choice and structuring writing, and revision. Readings and discussions included mention of diverse practitioners of science, how to interview scientists for journalism, what science is and how it's done, and reflection on existing perceptions of science. The course met once a week for 2 hours and 50 minutes for 10 weeks, with approximately 9 hours of out-of-class work each week for a total of 12 hours.

\subsection{Assessments and Data Collection}
Assessments were administered in Weeks 0 and 1 (pre-course) and in Week 10 and Finals Week (post-course) to gauge student attitudes before and after the course, in the form of a Google Forms survey. The survey contained both Likert-scale quantitative questions and open-ended written response questions. Student responses were not anonymous, but identifying information was only used to match pre- and post-survey responses. All protocols for this study were approved by UCLA's Institutional Review Board (IRB\#22-000440).

Likert-scale questions were drawn from the Astronomy and Science Student Attitudes (ASSA) instrument, recently developed to fill a gap in astronomy education research tools \citep{bartlett2018astronomy}, and the widely used Epistemological Beliefs Assessment for Physical Sciences (EBAPS) \citep{elby2001helping}. ASSA provides insight into discipline-specific attitudes across 8 factors, six of which are strongly relevant to this study: student interest in astronomy, interest in science outside of school, perception of ability in science, future aspirations in science, perception of benefits of science, and personal relevance of science. EBAPS captures broader attitudes about science as a whole across five dimensions: structure of scientific knowledge, nature of knowing and learning, real-life applicability, evolving knowledge in science, and source of ability to learn. Combined, these two instruments should provide a thorough view of students’ perceptions of science and their ability to engage with it. Additionally, some Likert-scale questions were designed for this course to specifically probe student attitudes towards science writing, using scales based on \citet{vagias2006likert}. However, this quantitative data is not analyzed beyond basic descriptive statistics in this work (see Section \ref{subsec:quant}); instead, we limit our scope to focus on mainly the qualitative data due to the small sample size.

\begin{figure}
    \centering
    \includegraphics[width=\linewidth]{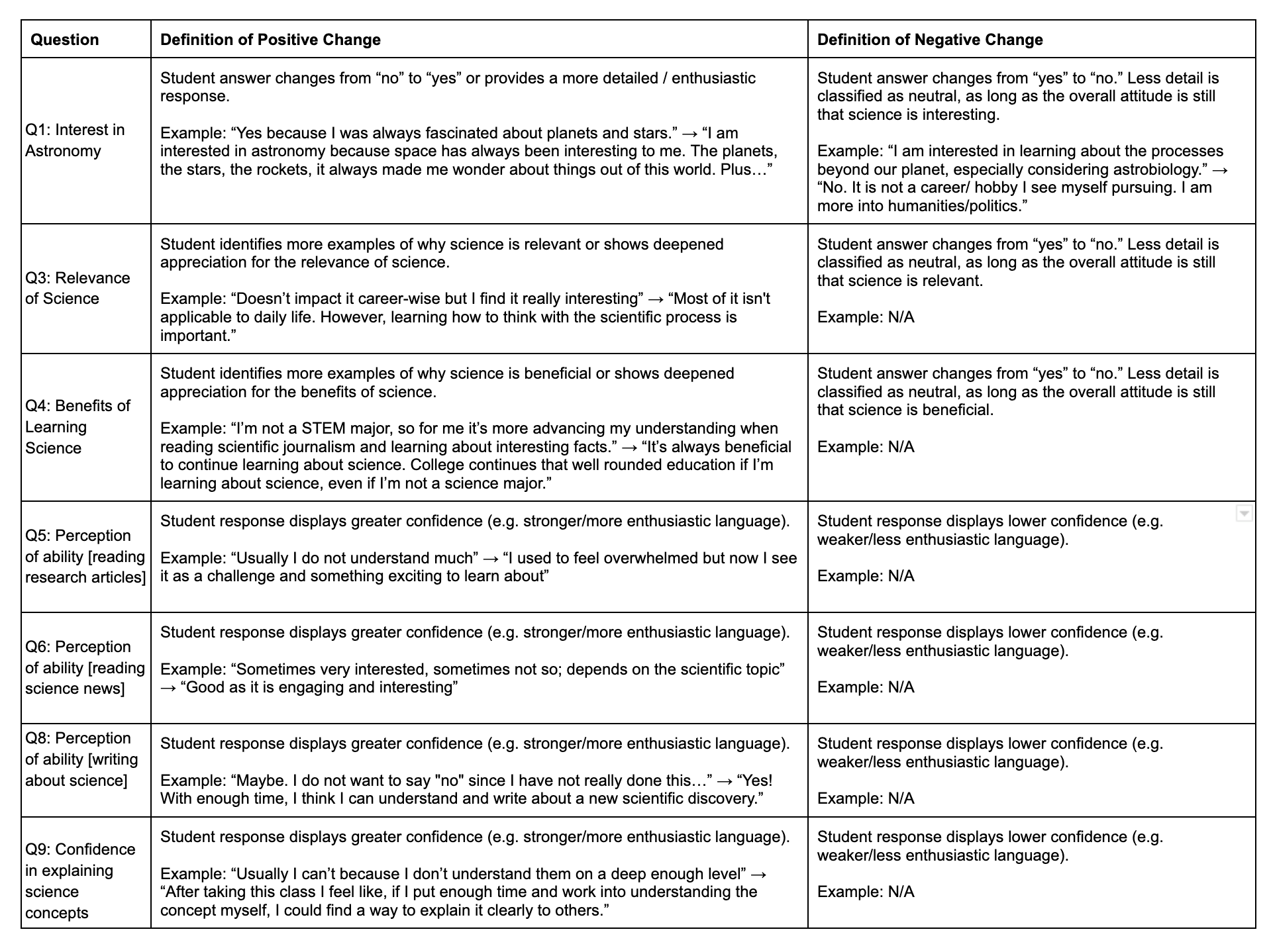}
    \caption{Definition of ``positive'' versus ``negative'' change for each question coded based on matched pre- and post-course responses.}
    \label{fig:change-coding}
\end{figure}

The open-ended questions are designed to gather nuanced information on student experiences and how they felt writing related to their changes in perspectives. Some of these questions were taken from \citet{raved2011attitudes}, which provided a qualitative look at student attitudes in a 10th grade classroom via interviews. The written response open-ended questions were the same on the pre- and post-course surveys, and were as follows:
\begin{enumerate}
    \item Are you interested in astronomy? Why or why not?
    \item Why do you want to learn about science journalism? 
    \item How is the science you learn in school relevant to your daily life? 
    \item How do you benefit from learning about science in college?
    \item How do you feel when reading scientific research articles?
    \item How do you feel when reading about science in magazines, newspapers, etc.? 
    \item What does it take to be a good scientist? 
    \item How do you feel about explaining science concepts to others? 
    \item Do you think you could write about a new scientific discovery? Why or why not? 
    \item Is it important for scientists to have good writing skills? Why or why not? \textit{(Note: this question is not used/analyzed in this work)}
    \item What do you think of the way science is portrayed to the public? \textit{(Note: this question is not used/analyzed in this work)}
\end{enumerate}
 Responses to Questions 1-9 were coded into various emergent categories, described in their respective sections below. Questions 1, 3, 4, 5, 6, 8, and 9 were also coded into 5 categories based on their change from pre- to post-course responses: positive change / improved, no change (positive), no change (neutral), no change (negative), and negative change / decreased as described in Figure \ref{fig:change-coding}. Positive change is defined as moving towards the desired outcome (e.g. increased interest in STEM, increased confidence and sense of ability in STEM, more diverse and open views of science and scientists). The three categories of ``no change'' indicate that student views have either remained positive, remained neutral, or remained negative. Student responses, particularly long quotes or those presenting multiple ideas, may be coded as multiple emergent themes, but each student responses is only counted in one of the change categories (e.g. an answer cannot count for both ``yes'' and ``no''). To visualize the results of the coding of student responses with respect to change in attitudes, we created a summary bar chart (Figure \ref{fig:bar}). 

The post-course survey included an option for students to be contacted for interviews about their experiences to gather further qualitative data. Student quotes from interviews and open-ended questions are reported in Section \ref{sec:results}, and the interview questions were as follows:
\begin{itemize}
    \item How has writing about science impacted your experience with learning about science?
    \item What do you feel is the most important lesson you’ve taken away from this course?
    \item Which of your views about science has changed the most while taking this course?
    \item How do you feel about engaging with science beyond this course?
\end{itemize}

Although this investigation had not yet been designed during the first offering of this course in Spring 2021, we did collect some pre- and post- course survey data as well as qualitative feedback from students about their attitudes towards writing and science and their experiences in the course. Students were asked for consent to use their responses to explore if there were any valuable insights that may be useful to this study. All student responses are from the 2022 study, unless otherwise noted as information pulled from the 2021 course evaluations.

\section{Results} \label{sec:results}

\begin{figure}
    \centering
    \includegraphics[width=0.8\linewidth]{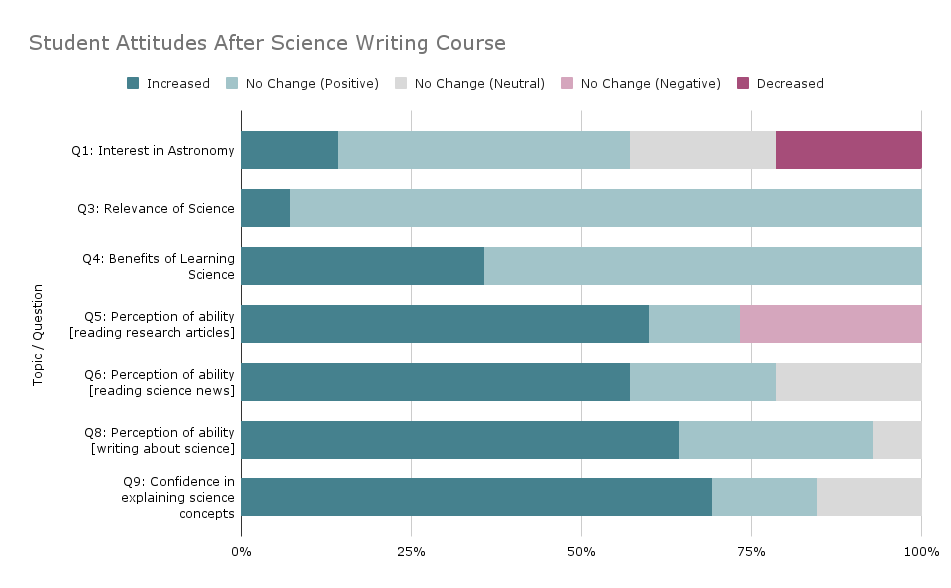}
    \caption{Bar chart illustrating changes in student responses from before and after the courses, for a sample of N=13-15 students. Dark teal indicates an improvement from students' starting opinions, and dark pink represents a more negative opinion (with negative and positive defined as in Figure \ref{fig:change-coding}). The three lighter colors -- light teal, grey, and light pink -- represent no change from the initial opinion, where students' opinions remained positive, neutral, or negative respectively. Students display little change in their interest in astronomy, but increases in their perceptions of the benefits of learning science, perceptions of their own ability to engage with science, and their confidence in explaining science concepts.}
    \label{fig:bar}
\end{figure}

A summary of changes in student attitudes based on responses to open-ended survey questions is presented in Figure \ref{fig:bar}. Results in this chart will be further discussed in the following subsections in five main categories: interest in astronomy, perceptions of benefits and relevance of science, perceptions of ability in science, perceptions of scientists, and other insights from interviews with student participants. We also briefly discuss the quantitative data gathered. Only responses from students who completed both pre- and post-course surveys were used to code for changes in attitudes.

\subsection{Interest in Astronomy}\label{sec:interest}

\begin{figure}
    \centering
    \includegraphics[width=0.6\linewidth]{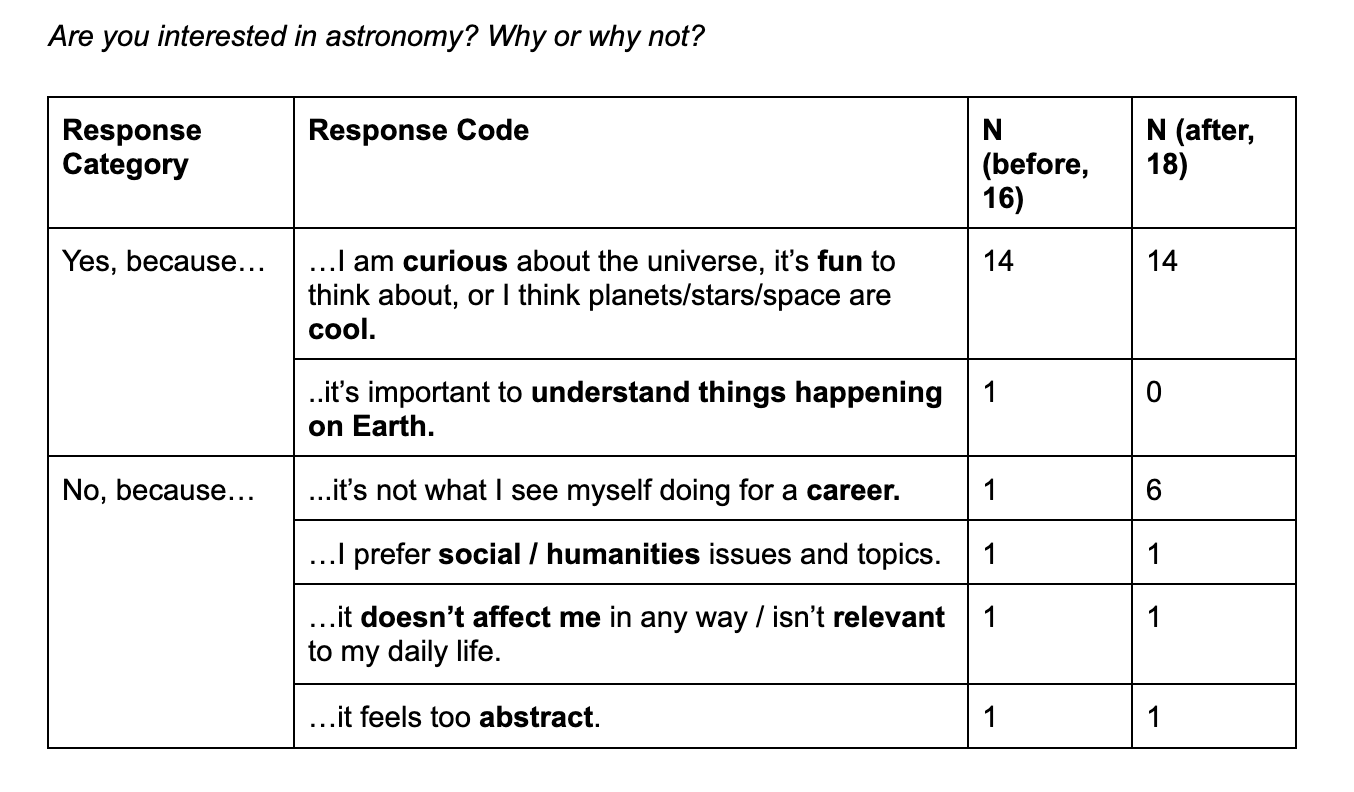}
    \caption{Student responses to the question ``Are you interested in astronomy? Why or why not?'' coded by the theme of their response. Students came in interested, and stayed interested, but confirmed their career choices in other fields.}
    \label{fig:interest-tab}
\end{figure}

Of all the metrics measured in this study, student interest shows the smallest positive change, and is the only metric to show a decrease in some students, as shown in Figure \ref{fig:bar}. Most students showed no change, either coming in with an interest in astronomy and maintaining that interest, or coming in neutral and remaining neutral. Narrative responses investigating student interest in astronomy were coded into two main positive and four negative categories, displayed in Figure \ref{fig:interest-tab}. Student responses expressing interest almost universally explained that their interest stemmed from a curiosity or sense of entertainment when thinking about their place in the Universe, while one student referenced astronomy as a way to understand what's happening on our own planet. Most of the decreases in claims of interest seem to stem from responses that students are not interested because it's not what they see as their career -- indicating that students may have interpreted the question somewhat differently than intended, as if ``interest'' requires a commitment to career on the topic. 

\begin{figure}
    \centering
    \includegraphics[width=0.6\linewidth]{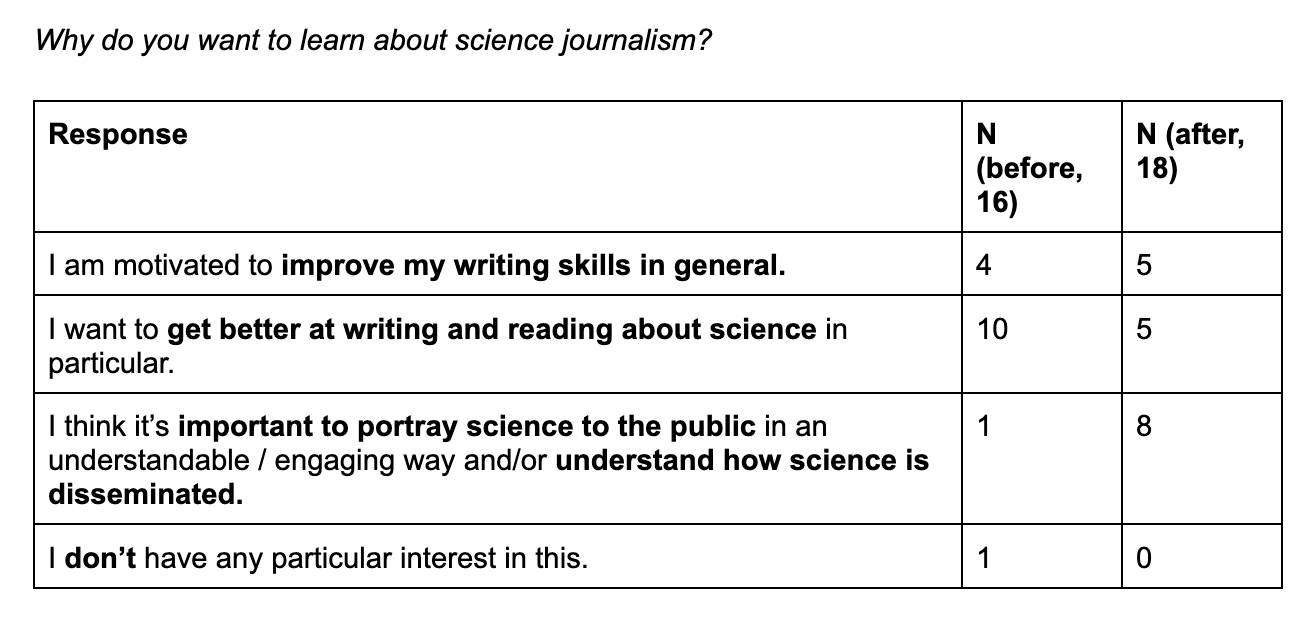}
    \caption{Student responses to the question ``Why are you interested in learning about science journalism?'' coded by the theme of their response. Interest level seems similar between before/after, but comments became more specific (e.g. mentioned things like jargon, audience, difference between research and public articles).}
    \label{fig:why-tab}
\end{figure}

As mentioned in Section \ref{sec:methods}, this student population likely had high interest coming in, due to the self-selecting nature of this course and their previous exposure to the topic in the earlier parts of the year-long program. Another question on the survey probed their interest in this course's other main topic -- science journalism -- upon entry to gauge reasons for taking the course. Students replied in four main categories: a desire to improve their general writing skills, a desire to improve their discipline-specific (e.g. science) writing skills, a belief in the importance of communicating science to the public, and a desire to better understand science by writing about it, summarized in Figure \ref{fig:why-tab}.

The following are examples of student responses conveying these sentiments, with their corresponding codes:
\begin{displayquote}
``It is interesting to see how different parts of life work, and that includes journalism. It is cool to feel like you are on the writing side for once when for the rest of your life, you are the one reading articles.'' [Code: Understanding portrayals/dissemination of science]
\end{displayquote}
\begin{displayquote}
``I want to learn about science journalism because it is important to understand how the science that we read about in publication is released to the public.'' [Code: Understanding portrayals/dissemination of science]
\end{displayquote}
\begin{displayquote}
“Science is important to understand even if it is not something you want to do with your life.” [From 2021 Evaluations]
\end{displayquote}
\begin{displayquote}
“Science is so cool and even though I might not understand all of the physics/math/computer coding, appreciating the topics conceptually is so important and I want to continue to stay informed.” [From 2021 Evaluations]
\end{displayquote}

\subsection{Perceptions of Benefits and Relevance of Science, and Scientists Themselves}\label{subsec:perc-field}

\begin{figure}
    \centering
    \includegraphics[width=0.6\linewidth]{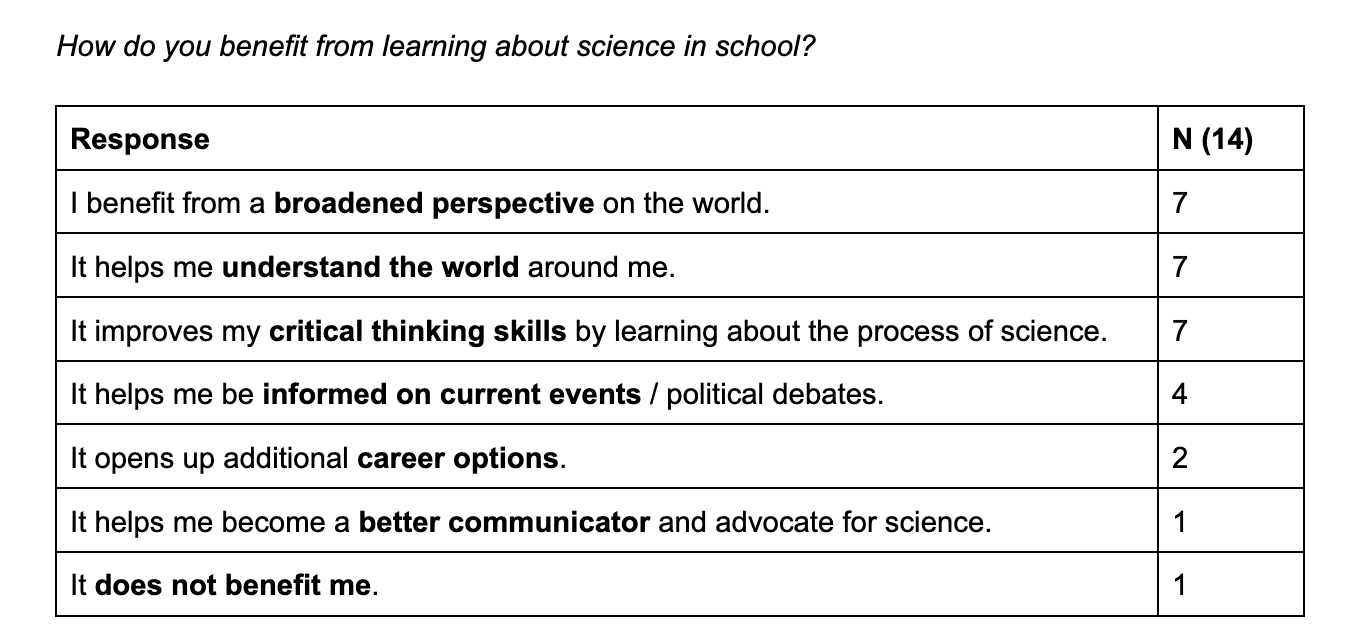}
    \caption{Student responses to the question ``How do you benefit from learning about science in school?'' coded by the theme of their response. Students had strong reasons for why science is relevant to their lives. 5/14 clearly show more positive perceptions of the benefits of learning science. All other students with responses for both surveys either showed an addition of nuance to their responses, or no change from views that were already positive.}
    \label{fig:benefit-tab}
\end{figure}

A majority of Americans currently think science is mostly beneficial to society, and perceptions of science as beneficial scale positively with science education level \citep{pewcenter_funk_2020}. Unsurprisingly, students nearly unanimously agreed that there was \textit{some} benefit to learning about science in college, either entering with positive views and remaining positive, or entering neutral and becoming more positive. (The benefits of science are emphasized in the first two quarters of the year-long program this course is a part of.) Their responses were coded into the following categories, listed in Figure \ref{fig:benefit-tab}: science broadens their perspective on the world, helps them understand the world, improves critical thinking skills, helps them be informed on current events, provides additional career options, and helps them become a better advocate for science. Only one student explicitly said that science did not benefit them, and this student did not fill out the post-course survey so it is unknown if the course changed their view.

\begin{figure}
    \centering
    \includegraphics[width=0.6\linewidth]{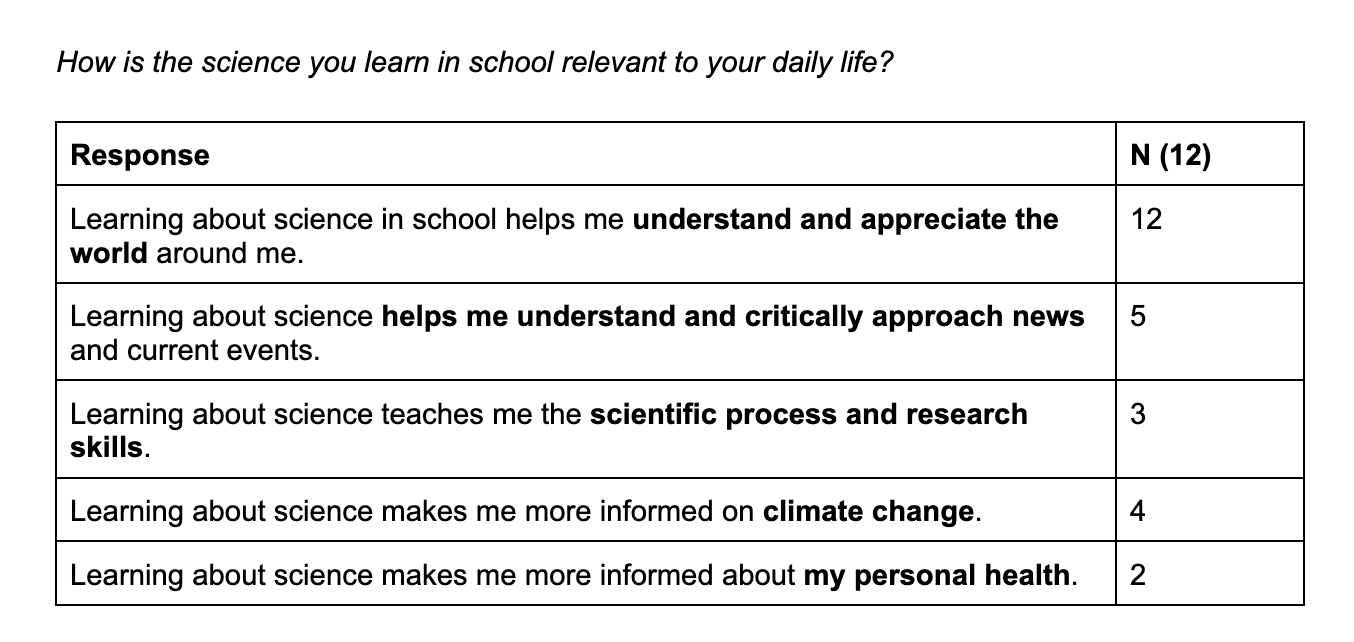}
    \caption{Student responses to the question ``How is the science you learn in school relevant to your daily life?'' coded by the theme of their response. Most students appreciated science coming in, but some added nuance or emphasis to their appreciation after the course.}
    \label{fig:relevant-tab}
\end{figure}

Similar to perceptions of the benefits of science, students already appreciate the relevance of science before the course, but added some nuance or detail to their responses after the course. Student responses fell into 5 main categories, which in turn parallel the categories of ``relevance for understanding scientific phenomena'' and ``relevance for being an effective citizen'' from \citet{stuckey2013meaning}. These categories of student responses, shown in Figure \ref{fig:relevant-tab}, are: relevant for understanding and appreciating the world, critically approaching news, understanding the scientific process, becoming informed on climate change, and being informed on personal health. These last two categories parallel some of the largest science-related events in the news at the time of this study: the climate crisis, and the COVID-19 pandemic.

\begin{figure}
    \centering
    \includegraphics[width=0.6\linewidth]{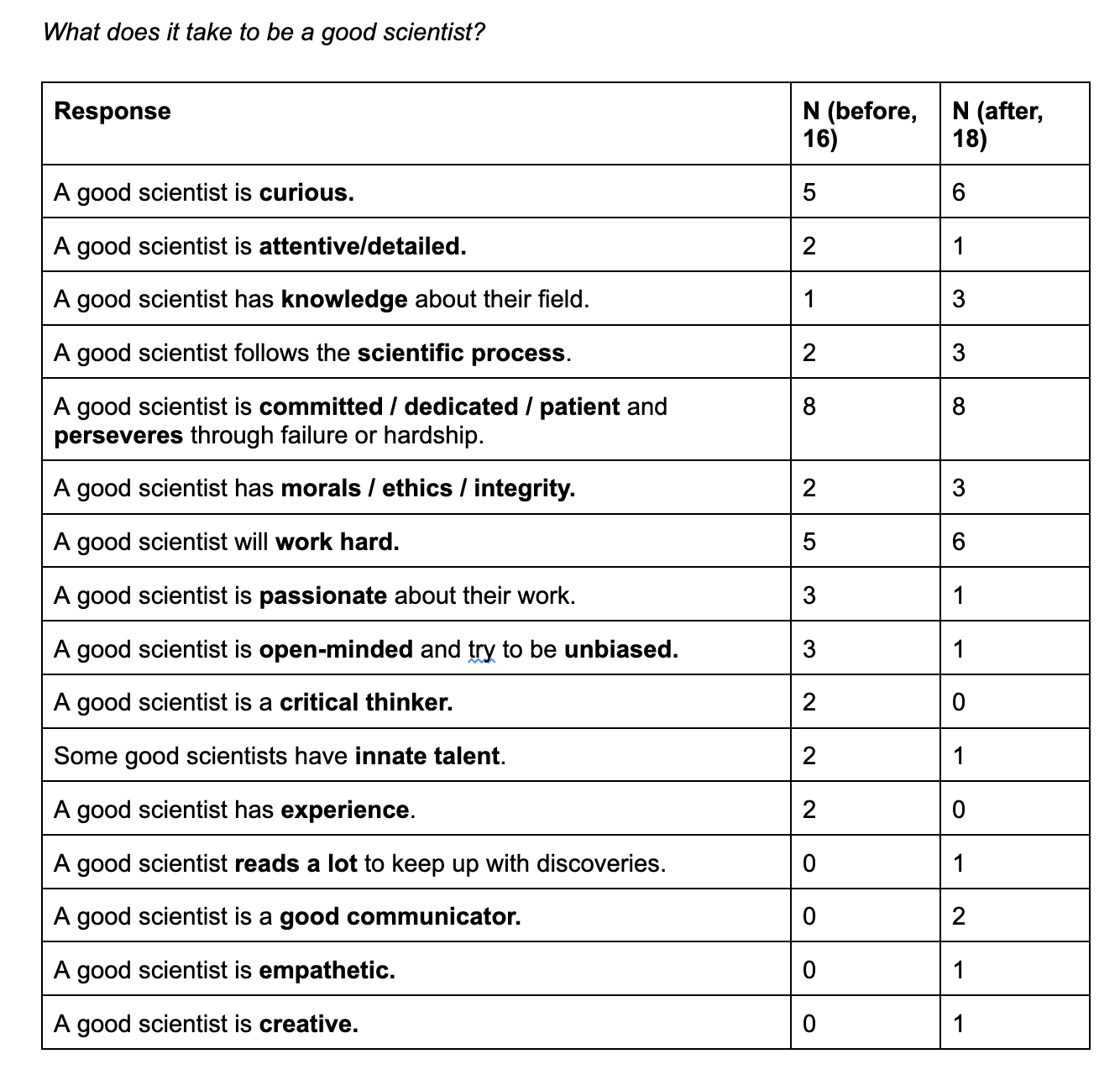}
    \caption{Coded student responses to the question ``What does it take to be a good scientist?'' Students list standard positive qualities with no strong notice of change before/after. Few students entered with stereotypes of requiring innate talent. There is a noticeable addition of mentions of ethics, morals, and communication to the required skills of a scientist post-course.}
    \label{fig:sci-change}
\end{figure}

Students in this course identified many of the expected positive qualities of a scientist -- hard-working, dedicated, curious, open-minded, passionate -- in both the pre- and post-course surveys, as visualized in Figure \ref{fig:sci-change}. Only 2/16 students responded with stereotypes requiring innate talent or solitary brilliance in the pre-course survey, and even then they mentioned it with a caveat (e.g. innate talent can help but hard work is needed). This is again unsurprising, as these are features of scientists that are explored in the first two quarters of the year-long program. In this seminar course, students were also able to make a direct connection to their instructor, an early career, female, first-generation astronomer, and the course featured representation from and discussion of other diverse scientists and writers. Additionally, awareness of negative stereotypes of scientists has existed for years \citep{losh2010stereotypes}, and it appears that the many mitigation strategies in place (e.g. outreach programs such as Skype a Scientist and pedagogy initiatives such as \citet{steinke1997portrait,thomas2006draw,yardley2015representation,jarreau2019using}) are making a difference. Interestingly, though, there was a noticeable addition of mentions of ethics, morals, and communication to the required skills of a good scientist after the writing-focused science course. One student left a particularly illustrative response on this point:

\begin{displayquote}
``My whole life I was told it was objectivity. Now I would disagree with that though. Science isn't objective, it is widespread, pervasive, and extremely personal. It is our future and our money. So, the most important qualities in science are being a good communicator and being ethical.'' [Codes: good communicator, morals / ethics / integrity]
\end{displayquote}

%\begin{figure}
    %\centering
    %\includegraphics[width=0.45\linewidth]{sci-pre.png}
    %\includegraphics[width=0.45\linewidth]{sci-post.png}
    %\caption{Word clouds generated from edited/coded responses to the question ``What does it take to be a good scientist?''. Students list standard positive qualities with no strong notice of change before/after. Few students entered with stereotypes of requiring innate talent. There is a noticeable addition of mentions of ethics, morals, and communication to the required skills of a scientist post-course.}
    %\label{fig:sci-change}
%\end{figure}

\subsection{Perceptions of Ability in Science}

Although students came in with positive perceptions of the importance of science as a field, and the belief that someone of any identity can be an effective scientist, these beliefs did not appear to extend to themselves. Many students entered extremely hesitant about their abilities to learn, write, and discuss about science topics. As shown in Figure \ref{fig:bar}, all four metrics of student ability to engage with STEM significantly increased after the course. We asked students about their feelings towards four tasks related to science: reading scientific research papers, reading popular science writing (e.g. magazines, news), explaining science concepts to others, and writing about a new scientific discovery.

\begin{figure}
    \centering
    \includegraphics[width=0.6\linewidth]{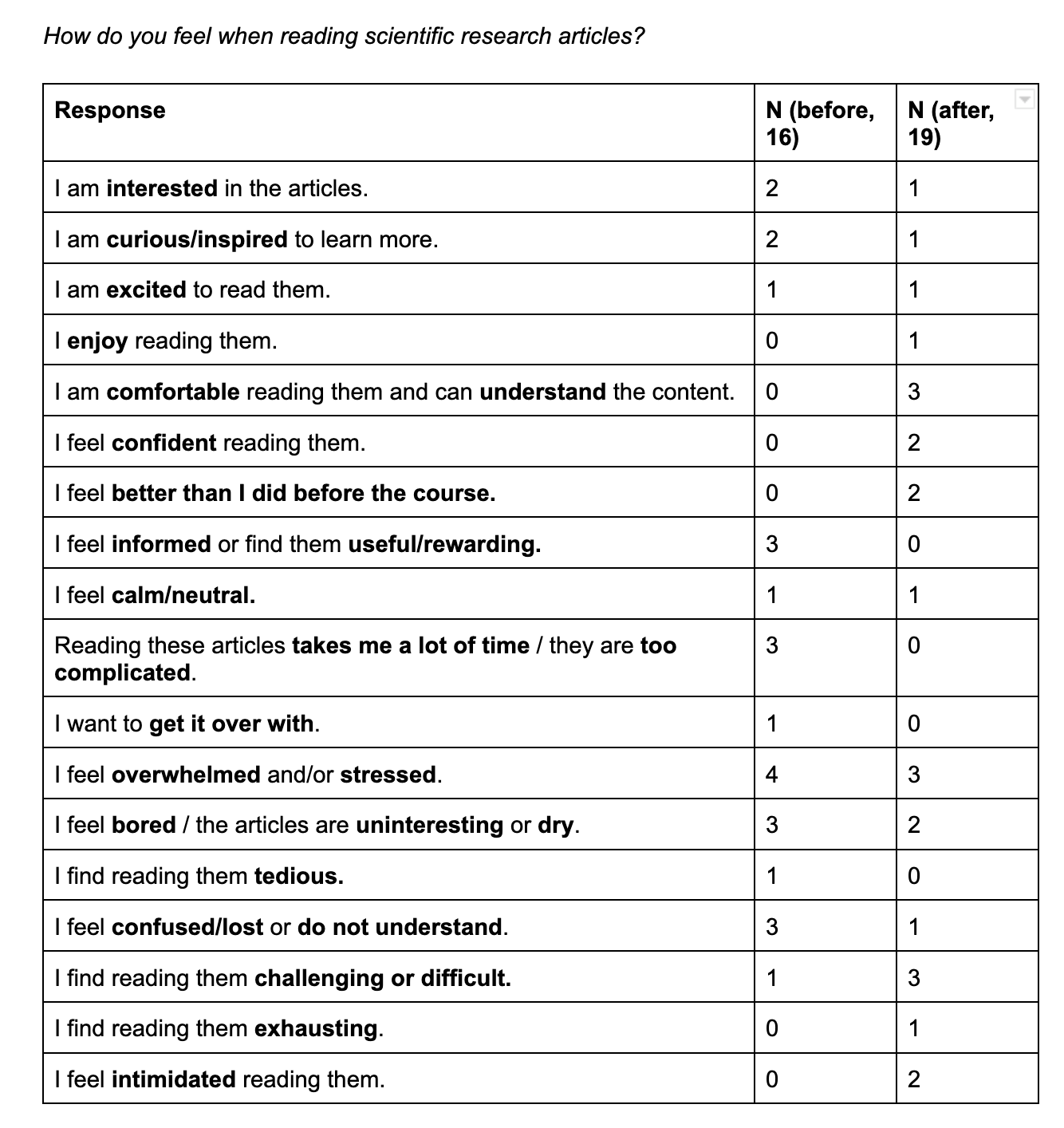}
    \caption{Student responses to the question “How do you feel when reading scientific research articles?'' coded by the theme of their response. There was a clear improvement in student comfort with reading primary research articles. 9/15 students report an improved attitude towards reading scientific articles. 4 report no change, and still find them intimidating, and 2 report no change, as they were already confident at the start.}
    \label{fig:research-feelings-tab}
\end{figure}

%\begin{figure}
    %\centering
    %\includegraphics[width=0.45\linewidth]{research-pre.png}
    %\includegraphics[width=0.45\linewidth]{research-post.png}
    %\caption{Word clouds generated from edited/coded responses to the question ``How do you feel when reading scientific research articles?''. Pre-course responses are on the left, and post-course on the right. }
    %\label{fig:research-change}
%\end{figure}

Students overwhelmingly began the course with negative attitudes towards the prospect of reading scientific technical writing, with ``overwhelmed'' being the most used word in their responses, as illustrated in Figure \ref{fig:research-feelings-tab}, while acknowledging that there may be interesting content hidden within. Although students were exposed to primary scientific literature in the other portions of the year-long program, those experiences do not seem to have made them feel more prepared for engaging with technical scientific writing -- it would be interesting in future work to determine if this prior experience changed student attitudes, determining if more typical non-STEM majors in a GE course hold the same beliefs. This metric shows an extremely positive change after the course, where students experienced directed reading assignments, guidance on how to tackle research articles, and assignments where they translated the big ideas of the articles for a more general audience. Students report feeling ``better'' ``comfortable'' and even ``confident'' in reading research papers post-course, with some even claiming to ``enjoy'' the challenge, while at the same time acknowledging that this is a difficult task. For example, see the following quotes from student responses:
\begin{displayquote}
``I used to be very intimidated by the jargon and the non-engaging writing style, but now I feel comfortable at least skimming those articles.'' [Code: comfortable]
\end{displayquote}
\begin{displayquote}
``I used to feel overwhelmed but now I see it as a challenge and something exciting to learn about.'' [Codes: challenging, excited]
\end{displayquote}

\begin{figure}
    \centering
    \includegraphics[width=0.6\linewidth]{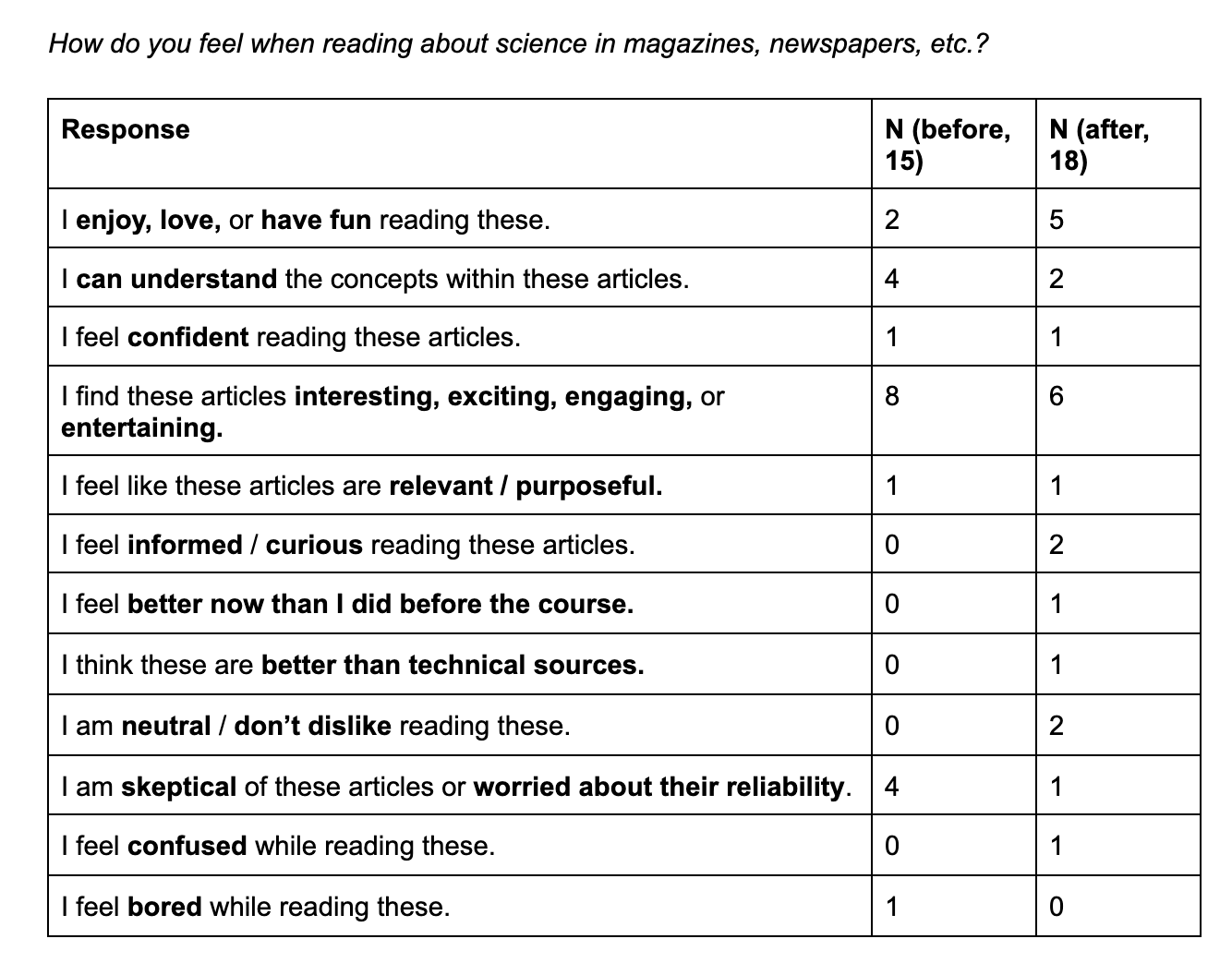}
    \caption{Student responses to the question ``How do you feel when reading about science in magazines, newspapers, etc.?'' coded by the theme of their response. 8/14 students report an improved attitude towards reading about science in the news. 3 show no change, remaining neutral about it, and 3 show no change, as they already enjoyed it coming in. No students like reading about science less than when they started. Responses also illustrated a greater understanding of genre, and the distinguishing features (namely audience) between research papers and popular science writing. Overall more positive attitudes both before and after than towards primary research articles.}
    \label{fig:pop-change}
\end{figure}

%\begin{figure}
    %\centering
    %\includegraphics[width=0.45\linewidth]{pop-pre.png}
    %\includegraphics[width=0.45\linewidth]{pop-post.png}
    %\caption{Word clouds generated from edited/coded responses to the question ``How do you feel when reading about science in magazines, newspapers, etc.?''. 8/14 students report an improved attitude towards reading about science in the news. 3 show no change, remaining neutral about it, and 3 show no change, as they already enjoyed it coming in. No students like reading about science less than when they started. Responses also illustrated a greater understanding of genre, and the distinguishing features (namely audience) between research papers and popular science writing. Overall more positive attitudes both before and after than towards primary research articles.}
    %\label{fig:pop-change}
%\end{figure}

In their daily lives, students are more likely to encounter secondary sources of science news, such as popular science articles in magazines. Unsurprisingly, given that students are the target audience of these types of publications unlike primary research articles, incoming attitudes about reading popular science articles were much more positive. Over half of students still report increased positive attitudes towards reading these types of articles, describing them as ``cool'' and ``interesting'' and saying they ``enjoyed'' them, as shown in Figure \ref{fig:pop-change}. Responses also illustrated a greater understanding of genre, and the distinguishing features (namely audience) between research papers and popular science writing. Students seemed to understand that these articles were \textit{meant} for them, and indicated appreciation for the form, such as in this student response:
\begin{displayquote}
``[I feel] much better than when I read scientific research. This material is accessible while educational, enabling someone without a professional background to understand the science involved. I feel like there's actually a point to reading it.'' [Codes: better than technical sources, purposeful, can understand]
\end{displayquote}

\begin{figure}
    \centering
    \includegraphics[width=0.6\linewidth]{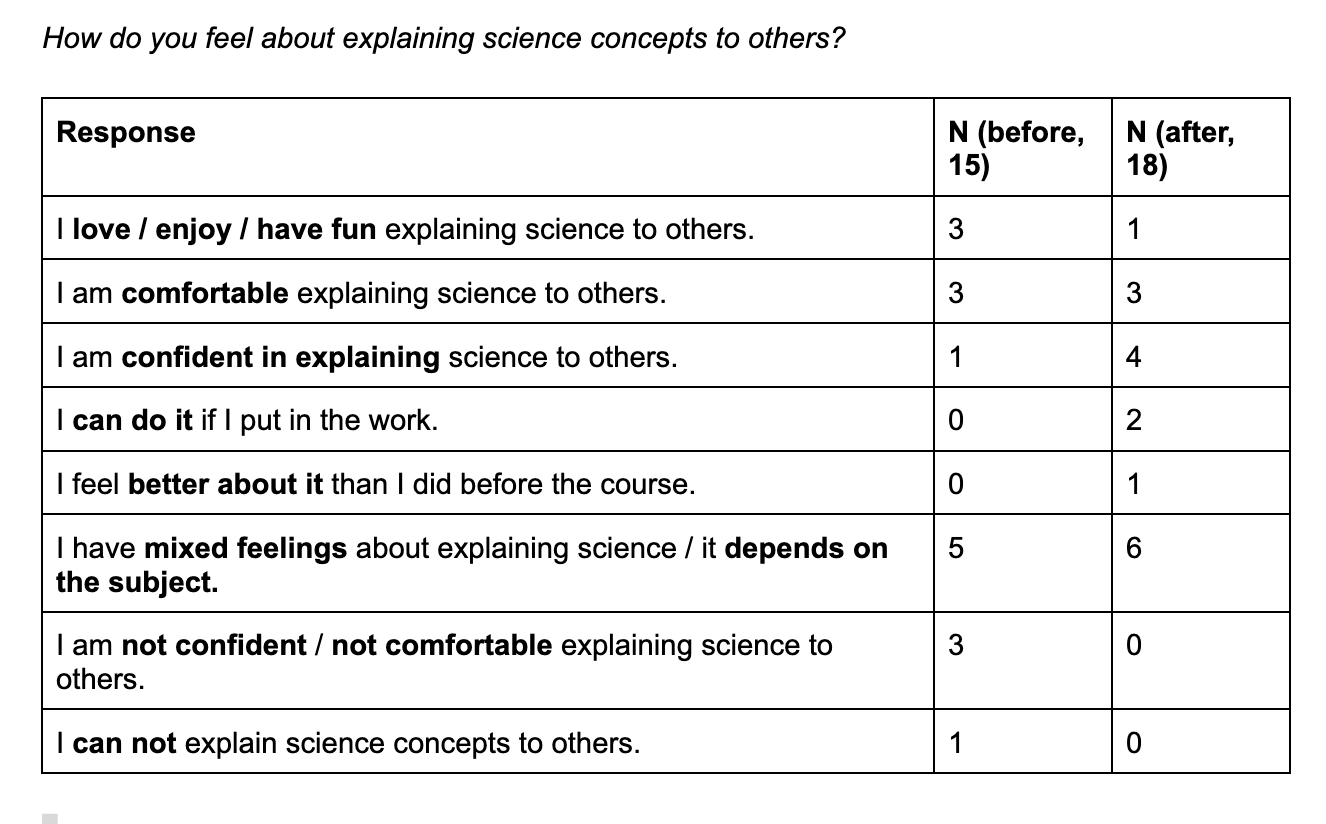}
    \caption{Student responses to the question ``How do you feel about explaining science concepts to others?'' coded by theme of their response. 9 students became more confident in explaining science, as defined in Figure \ref{fig:change-coding}, and 2 remained neutral while 2 entered already confident. The language students used in response to this question changed from comfortable (with mentions of worries of being incorrect), to a mostly confident.}
    \label{fig:con-change}
\end{figure}

%\begin{figure}
    %\centering
    %\includegraphics[width=0.45\linewidth]{con-pre.png}
    %\includegraphics[width=0.45\linewidth]{con-post.png}
    %\caption{Word clouds generated from edited/coded responses to the question ``How do you feel about explaining science concepts to others?''. 9 students became more confident in explaining science, and 2 remained neutral while 2 entered already confident. The language students used in response to this question changed from comfortable (with mentions of worries of being incorrect), to a mostly confident.}
    %\label{fig:con-change}
%\end{figure}

Beyond being a passive consumer of scientific information through reading, we also probed student abilities to \textit{actively} engage with science, such as through explaining scientific concepts to others. This is an extremely relevant skill to take away from an introductory science course, and can empower them to be scientifically literate and involved citizens. The language students used in response to this question changed from ``comfortable'' (with mentions of worries of being incorrect), to mostly ``confident,'' as shown in Figure \ref{fig:con-change}. Students also mentioned that practicing explanations helped them improve their own understanding, such as in the following responses:
\begin{displayquote}
``I think to be able to explain science concepts to others is a way to prove that you actually understand the concept yourself.'' [Code: N/A]
\end{displayquote}
\begin{displayquote}
``After taking this class I feel like, if I put enough time and work into understanding the concept myself, I could find a way to explain it clearly to others.'' [Code: can do it]
\end{displayquote}

\begin{figure}
    \centering
    \includegraphics[width=0.6\linewidth]{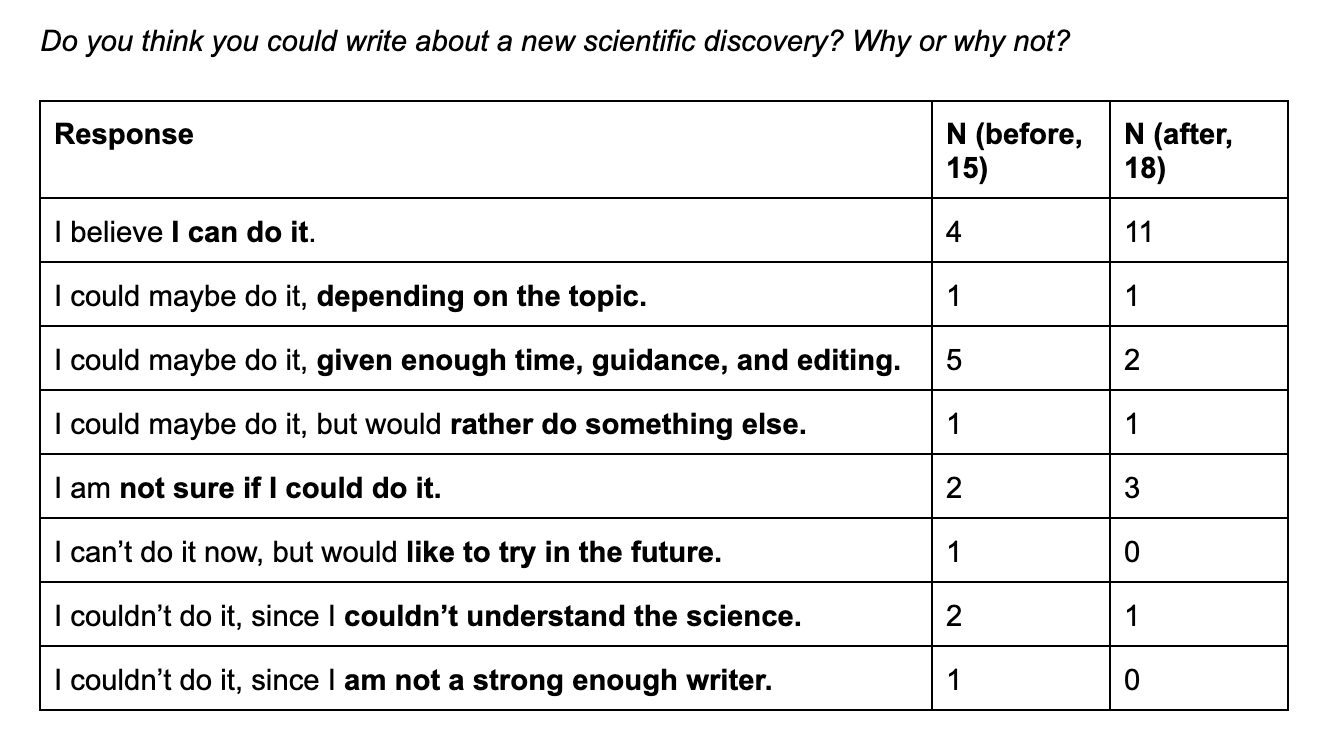}
    \caption{Student responses to the question ``Do you think you could write about a new scientific discovery? Why or why not?'' coded by the theme of their response. 9 students felt more confident about writing a new scientific discovery, 1 remained neutral, and 4 entered with confidence in their abilities. There is some evidence of additional understanding of different modes/audiences of scientific writing, and some students said their confidence level depended on the audience.}
    \label{fig:wri-change}
\end{figure}

%\begin{figure}
   % \centering
    %\includegraphics[width=0.45\linewidth]{wri-pre.png}
    %\includegraphics[width=0.45\linewidth]{wri-post.png}
    %\caption{Word clouds generated from edited/coded responses to the question ``Do you think you could write about a new scientific discovery? Why or why not?''. 9 students felt more confident about writing a new scientific discovery, 1 remained neutral, and 4 entered with confidence in their abilities. There is some evidence of additional understanding of different modes/audiences of scientific writing, and some students said their confidence level depended on the audience.}
    %\label{fig:wri-change}
%\end{figure}

Lastly, we gauged student attitudes towards writing about science before and after the course. This is arguably the most active and involved task related to engaging with science in this study, and requires students to synthesize many of the prior skills: reading and understanding primary sources, understanding the style, audience, and purpose of popular science writing, and explaining science concepts to a lay audience. Responses showed a clear positive trend, with students having more confidence in their abilities at the end of the course, as shown in Figure \ref{fig:wri-change}. Pre-course, students hedged with words and phrases like ``maybe'' and ``with help'' or ``with a good editor'' or ``with enough time.'' Post-course, students showed more confidence, but still wanted more practice and time to hone their skills. There is some evidence of additional understanding of different modes/audiences of scientific writing, and some students said their confidence level depended on the audience or topic. One student response really captures the transition from pre- to post- course attitudes and the value of a course focusing on science writing and communication:
\begin{displayquote}
``I think that, after taking this class, I could write about a new scientific discovery because I have been taught the process and worked through it with someone who has practice.'' [Code: can do it]
\end{displayquote}

% \usepackage{multirow}
% \usepackage{graphicx}
% \usepackage{booktabs}
% \usepackage{multirow}
% \usepackage{graphicx}

\subsection{Quantitative Data}\label{subsec:quant}

\begin{figure}
    \centering
    \includegraphics[width=0.5\linewidth]{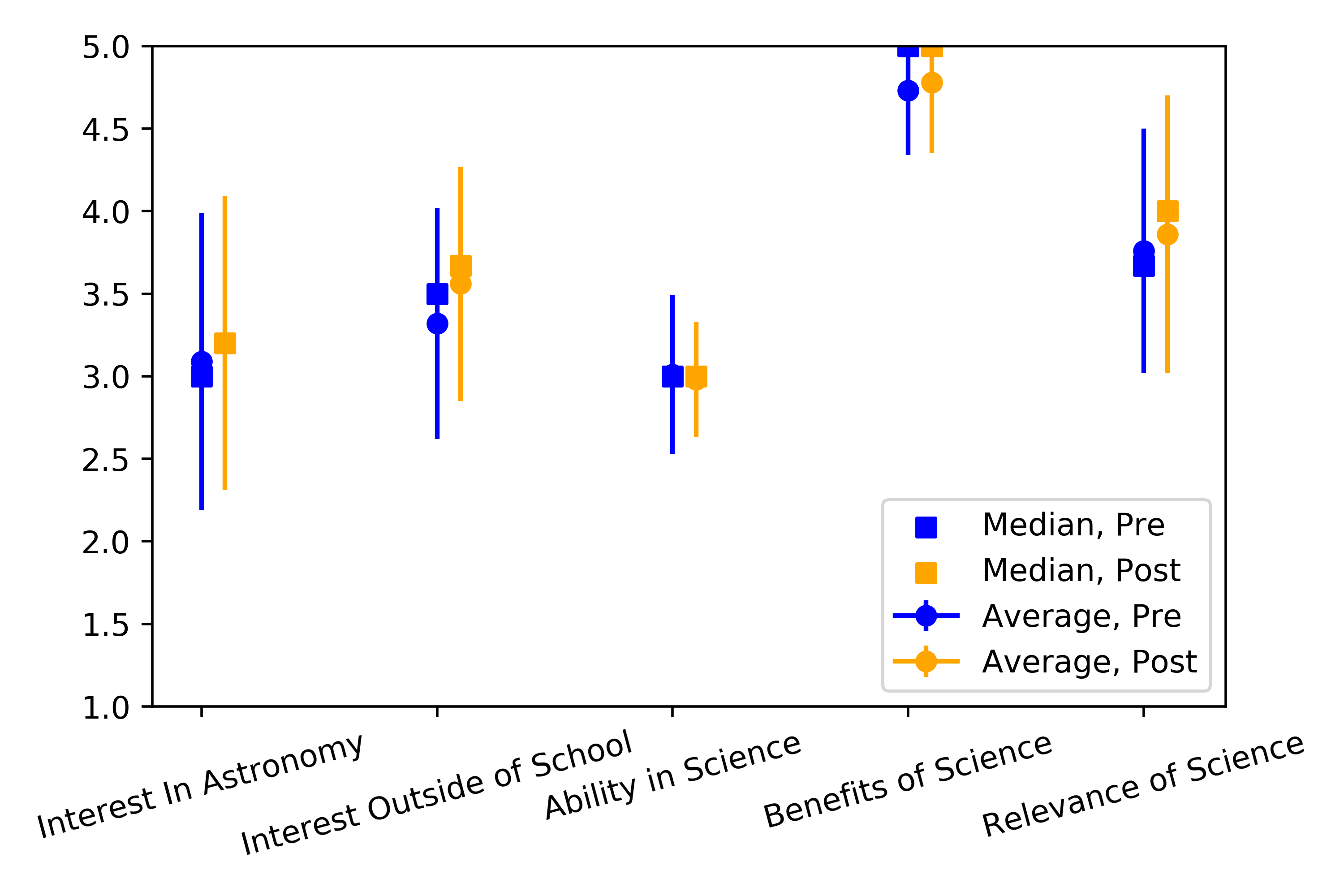}
    \includegraphics[width=\linewidth]{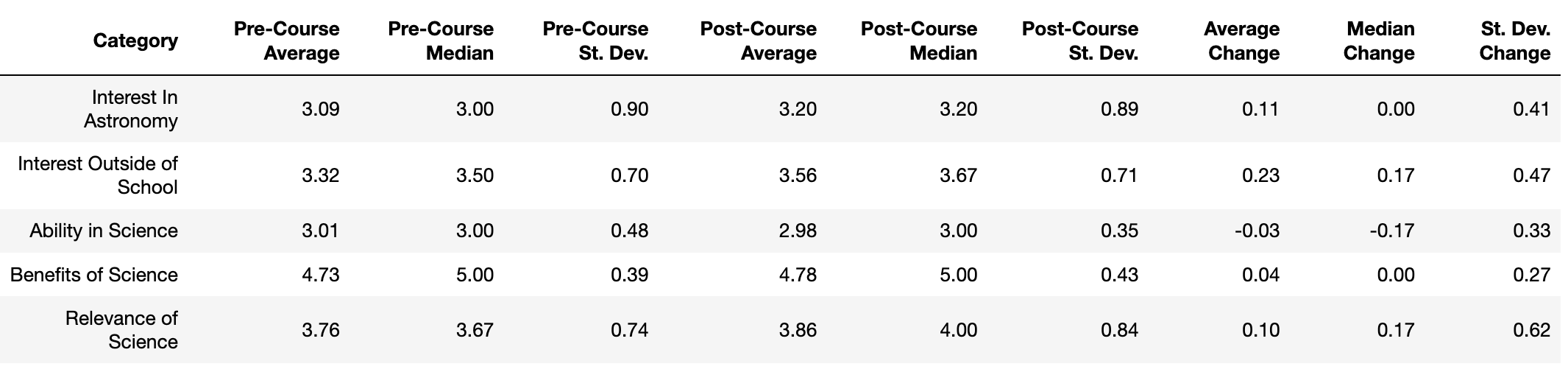}
    \caption{Summary of descriptive statistics from Likert-scale questions, combined based on their category, visualized above with the full data table on bottom. Improvements are seen in perceptions of relevance of science, interest in astronomy, and interest in general science, whereas perception of ability in STEM and benefits of science showed nearly zero change. P values are not listed, as none of them were significant at the p$<$0.05 level.}
    \label{fig:quant}
\end{figure}

Although the sample size is small, and accordingly the quantitative data does not show any trends significant at the p$<$0.05 level using a Wilcoxon signed rank test, we present descriptive statistics on the quantitative data to check for consistency with the qualitative data. These statistics are presented in Figure \ref{fig:quant}. The categories presented are based on the ASSA, described earlier in Section \ref{sec:methods} \citep{bartlett2018astronomy}. Category scores are calculated by taking the average of student responses over each question within the category.  It is interesting to see that the medians are so similar for ability in science and benefits of science (which was high to begin with), but there is an increase in interest and relevance of science. It is also worth noting that the questions for perception of ability from the ASSA were quite general, and may not have reflected student gains in specific skills and pathways for engagement with science covered in the course. For example, one of the ASSA items was that "Science is one of my best subjects." Given the students' strong preference and identity in the humanities, it is difficult to imagine them saying that science is one of their best subjects even after this course -- they all still knew, as evidenced in Figure \ref{fig:interest-tab}, that humanities was their bread and butter. We also hypothesize there may have been a mismatch in interpretation where they didn't think of this course as a ``typical'' science course, leading them to respond to items like ``I get good grades in science'' differently. In future studies, it may be worth using more specific items in any quantitative surveys.

\subsection{Insights from Interviews}\label{subsec:interview}

In this section, we present quotes from student responses to interview questions, and comment on their contents.

\subsubsection{How did writing about science impact your experience learning about science?}
\begin{displayquote}
“I think by having to relay the information in an accessible, clear manner, you have to really understand it…Like you have to understand it well, in order to explain it to different groups of people. So I think it just forced me to have a firm grasp of it, in order to be able to like, explain it clearly.”
\end{displayquote}

\begin{displayquote}
“With science, you really have to know what you're talking about. So if you weren't sure about something, it kind of inspires you to go research, go back and check.”
\end{displayquote}

\begin{displayquote}
“It really did. Because a lot of times science and STEM stuff feel super divorced from humanities. Which, if you're a humanities major, or just have humanities interests, like, oh, well, I guess this isn't really meant for me. But like, this is kind of marrying it. I love synthesis. And so this synthesis of like, the sort of critical thinking that you have to use in like, an English major, and kind of applying that to science is the best of both worlds for me.” 
\end{displayquote}

\begin{displayquote}
“Writing helped me really think about it from my perspective…when I was writing about it, I was able to actually think about my opinion on it. And, like, really get deep into it. So helped me like, synthesize what I learned, and form my view of the topic.”
\end{displayquote}

Students reported that the process of teaching material through written explanations increased their comprehension of the material, and the requirements to write and explain the material helped them to reiterate and remember the information in a way that they claim hadn't been present in their other STEM-related courses. Some students found that writing about science made the learning more ``manageable'' and others who had previously not enjoyed learning about science in the more ``traditional'' ways (e.g. from textbooks and lectures) found this method of active, project-based, writing-focused learning more interesting and engaging. Some students, including one quoted above, explained that incorporating writing into science education helped make science feel more relevant and related to their experiences and interests in the humanities, providing them with an entry point into the material based on their prior experiences -- more on this in Section \ref{disc:hum}. Others explained that writing about science forced them to consider their perspectives on debates and current events in science, and to dive deep into the topic at hand. This may connect to the earlier comment on objectivity in science in Section \ref{subsec:perc-field}, and a theme of students' view of the field of science expanding to consider ethics, communication, politics/opinion, and the role of individuals, which will be discussed further in Section \ref{disc:eth}.

\subsubsection{``What was your biggest takeaway from the course?''}
\begin{displayquote}
“We need to spread the word that scientists and like everybody needs to know how to write…the biggest thing I learned is probably that it's super important for the [science] writing to be accessible. That's what's gonna make people not be scared of the truth and the facts, is like if they can kind of understand it a little bit more.”
\end{displayquote}

\begin{displayquote}
“The most most important lesson is that I think things seem scary before you try it. Because all scientific writing, as I said before, was really intimidating. And it was just academic and boring for me. But after I, like, tried it, I mean, we had to kind of try it with their writing journals and stuff, after we started writing more often helped me realize that, like, things may seem scary at first, but when you try them, it's like, you'll get used to it and might not be as hard as you think it is.”
\end{displayquote}

\begin{displayquote}
“I think that a lot of classes should integrate more writing as part of like a science education…And I think I really would have benefited from that, just because that's part of the way that I learn.”
\end{displayquote}

Many of students' major takeaways from the course related to the elements of writing -- audience, style, structure, sentence-level concerns, etc. --  and their ability to connect different pieces of evidence into a coherent story, which is reasonable, and even expected or hoped for, given that this course fulfilled one of the students' major writing requirements. Other students mentioned takeaways about the science content, such as the idea that science is cool/awesome/interesting and their perspectives on life in the universe have changed (understandable again, given that this is a course focused on astrobiology and aliens.) Some students mentioned takeaways that related more to perceptions of science and the role of communication in STEM, such as the fact that they now see writing as an extremely important skill, even for scientists, and the idea that better science writing will lead to better science literacy in our society (see above quote). Others mentioned metacognitive takeaways, such as the idea that trying new things and pushing yourself is important for learning, as cited in the second quote of this section. They even directly identified writing as a key part of what made their learning successful in this course, claiming that more courses should integrate writing (as in the last quote above) and pinpointing that active learning activities like the writing in this course led to a better learning experience for them. We will discuss the takeaway of writing as active and experiential learning for science further in Section \ref{disc:exp}.

\subsubsection{``Which of your views about science changed most during the course?''}

\begin{displayquote}
``Probably the idea that it can be a creative endeavor as well...like, the scientific method, it seems very structured. And while I love structure, it's like...if you structure it too much, then that can feel confining.''
\end{displayquote}

\begin{displayquote}
``There are people who like...take articles and information and just make it accessible to people...There are people who are trying not to make it [science] as gate kept.''
\end{displayquote}

\begin{displayquote}
``Scientists need to communicate, because all of them are doing different things, and it's very naive to believe that they're all on top of each other's work all the time, even though that's kind of the impression you get from not being in that field.''
\end{displayquote}

Students report leaving this course with a better understanding of the media landscape surrounding science, likely due to the unique focus on science journalism in this course. As a result, this also changed their views of science itself -- one student cites a change in his previous view of science as a centralized ``institution,'' now understanding that science is more of a collaborative collection of people working independently but cooperatively. Another, quoted above, mentions their expanded view of \textit{who} participates in science and how -- they were previously unaware of writing and outreach roles in science, jobs beyond the stereotypical scientist role, which they were excited to learn existed to make the field more accessible to people like themselves. This student described this expanded view as helping them move away from the idea that science is ``stuffy'' or ``elitist.'' Similarly, students gained a greater appreciation and more accurate picture of what being a scientist is actually like on a day-to-day basis, upending tropes of boring scientists, becoming familiar with the activities of the job (e.g. analyzing data, communicating results), and understanding that science requires creativity, not just following a rigid set of rules. One student expressed a distaste with the idea of science as rigid, and almost a sense of relief at the realization that there was a connection between the creativity they sensed in their own humanities and artistic studies and real science. Additionally, other students expressed an appreciation that the skills they learn in humanities courses (e.g. writing, communication) actually \textit{are} relevant and transferrable to science, helping to bridge the often daunting gap between the two fields (discussed further in Section \ref{disc:hum}). Another common thread amongst student responses was that this course changed their view on whether or not life exists beyond Earth -- an interesting outcome, and somewhat expected for an astrobiology course, but not particularly relevant to this investigation.

\subsubsection{``How do you now feel about engaging with science, and has that changed since the beginning of the course?''}\label{q:engagement}

\begin{displayquote}
“I can approach it [science] with more understanding, like if I'm reading an article or something, or like listening to a new story about whatever. I feel like I can grasp the full picture. Like when they launched the telescope [JWST], I feel like I understand like, okay, like, this is what it's like trying to achieve and do, even if they don't say that explicitly.”
\end{displayquote}

\begin{displayquote}
“I knew going into college, I didn't want to like be in STEM, I didn't want to necessarily be practicing science. But that doesn't mean that I can't like engage with it. So that's kind of like my big takeaway. Like, I could still for fun, like write science articles online if I wanted to, or just things like that. I didn't really ever think about that [before].”
\end{displayquote}

In the interviews, students expressed not only an increased interest and confidence in engaging with science, but shared some concrete ways in which they are taking steps to engage more with science in their lives. One student decided to take some climate change courses as part of their undergraduate education, and one is considering adding a science minor after their experiences in this course. Another student, involved in a journalism group on campus, mentioned that they now have more confidence to take on journalism assignments related to science, whereas before this course they would pass on science-related assignments offered by their editors. 

As illustrated in the first quote of this section, students also mentioned that this course and the science writing instruction within helped them know what to look for in science news, providing them with both reading skills and science content to help determine if an article is reliable or worth their time to read. Clearly, this course impacted their confidence in their ability to be a critical consumer of science-related news and journalism -- a key skill for being an informed and active modern citizen.

Additionally, the second quote of this section highlights an interesting theme -- that someone who is interpreting, but not necessarily synthesizing or creating new scientific information, can have an identity within science as a field. This is discussed further in Section \ref{disc:con}.

\subsubsection{``Is there anything else about how this course impacted you that you’d like to share?''}
\begin{displayquote}
“Your course just kind of broadened even more like the idea that anybody can be a scientist, you just have to kind of try at it and be curious. And like anybody can learn things. Definitely it's helped me move on to be a little less nervous about jumping into things that I'm not entirely sure about, like, scientifically.”
\end{displayquote}
\begin{displayquote}
“I can write about science for the rest of my life without even having to formulate my own experiments/do research! (Not that I have anything against research, I was just surprised by how accessible the process of science writing actually is.).” [From 2021 Evaluations]
\end{displayquote} 

Lastly, space was given to students to mention anything else relevant to their course experience. Many students commented on facets of the course, such as their thoughts on the course instructor, which were not relevant to this investigation -- however, there were two relevant and interesting responses from students, which are quoted above. One explicitly states ``that anybody can be a scientist'' -- an oft-quoted goal for student conceptions of scientists in our increasingly more diverse and inclusive field \citep{rahm2002scientist,rubey2022anyone}. Additionally, the second student response here further illustrates the broadened conception of participation in STEM, as discussed in the preceding section.

\section{Discussion} \label{sec:disc}

Student feedback and responses from this investigation revealed a number of interesting takeaways for both \textit{how} and \textit{why} science writing education was impactful for their learning and views surrounding STEM. In this section, we discuss two major themes of how this learning experience affected their views, and two major themes of why this experience was so influential and science writing was a useful tool for learning about astronomy / science.

\subsection{How: Expanded Views and Increased Confidence in Ability for Engagement with Science}\label{disc:con}

\textbf{Takeaway \# 1:} Writing-focused science education expands how students think of engaging with science, providing new pathways beyond careers for them to be involved in STEM, and greatly increases their confidence in their own ability to engage with STEM in various ways.

In the student responses displayed above in Section \ref{sec:results}, students showed increased confidence in a variety of tasks, covering a range of cognitive processes and domains of knowledge \citep{bloom1956handbook,krathwohl2002revision,anderson2001taxonomy}, related to engagement with science content: 
\begin{itemize}
    \item \textbf{Remember}, \textbf{Understand}, and \textbf{Evaluate}: Critically reading articles about science, for both general and technical audiences
    \item \textbf{Remember}, \textbf{Understand}, and \textbf{Apply}: Explain science concepts to others
    \item \textbf{Understand}, \textbf{Analyze}, and \textbf{Create}: Write about a new scientific discovery
\end{itemize}

Additionally, students, both in survey responses and interviews, demonstrated a broader understanding of how people can be involved with science -- an expansion of students’ prior conception that only active researchers, who pursue STEM as a career, can engage meaningfully with scientific knowledge. Often, the conceptualization of what it means to be ``involved'' with science is extremely limited to those who are active practitioners within professionalized scientific careers (that is, those practicing science as their main job), and students are aware of few in-roads to further engagement with science beyond science as a profession \citep{avraamidou2021aspires}. The quotes in Section \ref{q:engagement} illustrate that this course, including exposure to science writing and more explicit discussions about how science works, may have been able to broaden student knowledge and awareness of varying levels of engagement with STEM (e.g. as a career, citizen science projects, amateur astronomy, writing and journalism, consuming news) that are available as valid options in their lives. Students seem to see engagement with science, from the citizen/non-professional point-of-view, as a meaningful part of science -- an attitude that science communication and citizen science efforts try to foster \citep{bonney2016can,williams2000getting,azevedo2018seeing,ibrahim2015student}.

Determinations of the level of the change in student engagement (e.g. no interest vs. appreciation of news, passing interest vs. career) are currently limited by the data collected, but could be probed in future work. It will be important to consider what our goals are for students' engagement levels, since we can not (and should not) aim to convert everyone to a STEM major or career; it is also worth articulating exactly what we hope to gain with increased engagement, and this should be considered in future discussions and investigations.

\subsection{How: The Roles of Ethics, Communication, Individuals, and Opinion in Science}\label{disc:eth}

\textbf{Takeaway \# 2:} Writing-focused science education creates students who are better prepared to deal with science in a real-world context by emphasizing important “soft” skills used both in research and research-adjacent outreach work. By discussing ethics, effective communication, and the agency available to individual scientists, students gain a more comprehensive understanding of work in and around science and develop new ideas about the skills and norms of an ideal scientist.%Writing-focused science education provides a clear pathway to illustrate the role of ``soft'' skills in science, such as the need for ethics and effective communication, and discuss the structure of scientific fields (including research-adjacent jobs, like outreach) and the agency of individual scientists, creating students who are better-prepared to engage with science in real-world contexts.

In this course, students were exposed to a variety of careers related to science beyond the pure research scientist jobs they are used to seeing represented (e.g. teaching, outreach, writing, policy, etc.). Within the scientific community, communication is increasingly well-established as a critical skill for scientists \citep{fleming2009talking}, but even in studies of student perceptions of STEM jobs, these roles are not taken into consideration \citep{christensen2014student}. After this course, students reported an expanded view of the different skillsets and roles it takes to make ``science'' happen, and they noticed the importance of communication -- both within scientific communities and between scientists and the general public. Another interesting observation by a student was that they no longer saw science as a monolithic institution, but instead understood that there are individual people at work -- all with their own opinions and even goals -- working collaboratively.

Students also didn't list ethics or communication as traits of science/scientists before the class, but they did after -- this is an interesting change, and additional evidence of students building their understanding of science in context and recognizing that science is political \citep{brown2004reason, ball2021science}, which is crucially important. For students to be able to critically approach science as it appears in their lives and become active and informed citizens, they must gain a greater understanding of the human elements of science, including ethics, politics, and science communication \citep{backhaus2019acknowledging, hodson2003time, elgin2011science,feinstein2013outside}. Some programs are already incorporating this sort of discussion explicitly in Science-Technology-Society (STS) programs \citep{han2014improving}, such as the Human Biology and Society undergraduate major at UCLA \citep{humbiosoc} and the Science Technology and Society Studies graduate certificate at the University of Washington \citep{uwstss}, but this investigation shows that science-writing-focused pedagogy can provide a platform for exploring these topics through discussion. Discussions of the role of science communication in science and public discourse were provoked by lessons on audience (who we're writing for in different genres, e.g. research papers vs. magazine articles), whereas conversations on ethics (for science journalists, scientists, and other stakeholders) and politics came about when reading recent science news articles critically as examples of good writing.

\subsection{Why: Science Writing as Active and Experiential Learning}\label{disc:exp}

\textbf{Takeaway \# 3:} Writing-focused science education enhances learning because it is active and experiential, immersing students in a real-world task and requiring significant and meaningful engagement.

Active learning strategies -- wherein students must actively engage in tasks related to their learning, instead of passively receiving information -- have been repeatedly proven to be beneficial to student learning and concept mastery \citep{ritchhart2011making,zayapragassarazan2012active,chickering1987seven,prince2004does}. Writing is an extremely active form of learning, requiring research skills, synthesis, and self-directed inquiry \citep{hamdan2005nonlinear}, and writing exercises have been shown to be an effective strategy for active learning \citep{linton2014identifying,butler2001active}. Similarly, peer learning has been demonstrated as an effective tool for increasing student learning \citep{hamilton2021peer}, and writing-to-learn is an established strategy in writing pedagogy \citep{ackerman1993promise,rivard1994review,bangert2004effects}. The use of writing, both in writing-to-learn activities and more polished journalism activities, in this course seems to have positively affected student learning of the content as well. 

Additionally, journalism / science writing assignments are relevant, real-world tasks that connect to aspirational goals (e.g. publishing an article) beyond the course; real-world assignments and incorporating recent discoveries are both effective strategies for increasing student engagement in the classroom \citep{nieves2020bitescis,eales2016establishing,laware2004real,wollschleger2019making}. Reading news articles critically is also a real-world task, one that we hope students carry through their lives -- this is the crux of science literacy, and especially by talking about current events, they know that their classwork is relevant.

\subsection{Why: Science Writing as a ``Relatable'' Entryway for Humanities Students}\label{disc:hum}

\textbf{Takeaway \# 4:} Writing-focused science education enhances learning because it relates otherwise seemingly irrelevant/unapproachable material to the primary work of a non-STEM major / someone who does not wish to pursue a career in science.

There are a number of well-known challenges in introductory STEM courses. One of these challenges is that humanities students sometimes see science as just fulfilling a requirement, and are therefore not that invested. Increasing student buy-in \citep{cavanagh2016student,wang2021framework} and demonstrating the relevance \citep{stuckey2013meaning, newton1988relevance,frymier1995s} of the content are key issues in motivating students in class. Student comments on their perceptions of the relevance of science to their lives are discussed earlier in Section \ref{subsec:perc-field}, but relevance of \textit{course content} is a different issue than their absolute interpretation of the relevance of science. Even science content, otherwise possibly irrelevant to their lives, can be made relevant with effective teaching techniques, such as incorporating current events and research. This investigation shows that science-related writing instruction may be another avenue to increase relevance; students report feeling that they are still gaining valuable skills in reading, writing, and critically interpreting science-related current events, even if they are learning content that is not directly related to their careers or aspirations.

Additionally, since science is often framed as opposition to humanities, there exists an aura of fear/intimidation around STEM fields for many who see themselves as ``non-scientists.'' By incorporating writing -- something students are familiar with -- into science courses, we can reduce that barrier to entry. Student familiarity to some part of the task at hand can improve learning outcomes and motivation \citep{soppe2005influence}. In this study, students do in fact report feeling that this course was more ``relatable'' or ``approachable'' than other STEM courses, indicating that the approach may be beneficial for introductory science courses and may make a larger or more lasting impact on student learning of science content and especially their attitudes towards science.

\section{Conclusions} \label{sec:conc}

After a quarter-long writing-intensive introductory science course, the culmination of a year-long GE course, student attitudes towards astronomy improved in multiple ways -- an increased perception of the benefits of science, a more nuanced and modern view of science as a field and scientists as people, and greatly increased confidence and belief in their own ability to engage with science. In interviews, students expressed directly that the science writing instruction was critical to their engagement and learning, essentially forcing them to develop a greater understanding of the material and helping them find purpose in course content through tasks that were both relevant to the real-world (e.g. interpreting science news) and relevant to their humanities-focused majors and prospective career paths. 

Overall, this qualitative study explored the effects of science writing on non-STEM major introductory students, and found that there were many positive impacts, adding evidence to the idea that interdisciplinary pedagogy including a focus on writing skills is useful for not only science majors, but also students learning about science who do \textit{not} plan to pursue a STEM career.  This is a first step towards gathering evidence for exploring and expanding writing pedagogy in science coursework, particularly in physics and astronomy where communication training is often absent. This course helped foster many desirable attitudes in students which point towards increased science literacy and breaking down stereotypes of scientists, and the following investigation provided evidence that discipline-based writing pedagogy in STEM courses has benefits and should be explored further with research and pedagogical interventions.

\begin{acknowledgments}
This material is based upon work supported by the National Science Foundation Graduate Research Fellowship under Grant No. 2021-25 DGE-2034835. Any opinions, findings, and conclusions or recommendations expressed in this material are those of the authors(s) and do not necessarily reflect the views of the National Science Foundation. This study was conducted with a UCLA IRB Exemption Certificate 	IRB\#22-000440.

Thank you to UCLA CIRTL for your support and guidance throughout this ``Teaching as Research'' project. Thanks to Sophie Bartlett and David McKinnon for providing further information on the ASSA and its interpretations. B. Lewis would also like to thank Laurel Westrup, Peggy Davis, Liz Galvin, and Greg Rubinson for their mentorship and motivation to study science writing in her discipline, and thanks to Casey Shapiro for guidance with regards to IRB approval.
%Graham Read, Katie Dixie, and K. Supriya for their mentorship in CIRTL's Teaching as Research Program; Tony Friscia for serving as course faculty advisor for Cluster 70CW; and Rachel Kennison for serving as faculty sponsor for this IRB process. 
\end{acknowledgments}


\vspace{5mm}





\bibliography{sciwri,atc}
\bibliographystyle{aasjournal}



\end{document}